\documentclass[epj]{svjour}
\usepackage{graphics}
\usepackage{color}
\usepackage{colortbl}

\begin{document}
\title{Rheological properties vs Local Dynamics in model disordered materials at Low Temperature}
\author{C. Fusco\inst{1,2} \and T. Albaret\inst{1} \and A. Tanguy\inst{1}}% etc
\offprints{ Anne.Tanguy@univ-lyon1.fr}          % Insert a name or remove this line
\institute{ \inst{1} Institut Lumi\`ere Mati\`ere, UMR5306 Universit\'e Lyon 1-CNRS, Universit\'e de Lyon, F-69622 Villeurbanne Cedex, France
\inst{2} INSA-Lyon, MATEIS CNRS UMR5510, F-69621 Villeurbanne Cedex, France}
\date{Received: date / Revised version: date}
\abstract{
We study the rheological response at low temperature of a sheared model disordered material as a function of the bond rigidity. We find that the flow curves follow a Herschel-Bulkley law, whatever is the bond rigidity, with an exponent close to $0.5$. Interestingly, the apparent viscosity can be related to a single relevant time scale $t_{rel}$, suggesting a strong connection between the local dynamics and the global mechanical behaviour. We propose a model based on the competition between the nucleation and the avalanche-like propagation of spatial strain heterogeneities. This model can explain the Herschel-Bulkley exponent on the basis of the size dependence of the heterogeneities on the shear rate.
\PACS{{83.50.-v}{Rheology. Deformation and flow.} \and
      {71.55.Jv}{Disordered structures, amorphous and glassy solids.} \and
      {47.57.Qk}{Complex fluids. Rheological aspects.} 
     } % end of PACS codes
} %end of abstract
\maketitle
\section{Introduction}
\label{intro}
Many disordered materials, such as glasses, foams, colloidal suspensions and granular matter exhibit a strongly heterogeneous mechanical response when submitted to an external driving~\cite{b.bailey,b.besseling,b.berthier1,b.berthier2,b.mobius,b.schall1,b.gibaud,b.dacruz}. This heterogenous macrosopic mechanical response, typical of an amorphous system, is a signature of local dynamical heterogeneities which appear in the form of collective erratic localized rearrangements, as shown both experimentally~\cite{b.besseling,b.gibaud,b.dennin,b.majmudar} and theoretically ~\cite{b.varnik,b.shi,b.tanguy,b.tsamados,b.fusco,b.delogu}. All these different materials are dominated by yielding properties, i. e. they behave like weak elastic solids at low stresses whereas they flow like viscous liquids above the so-called yield stress~\cite{b.divoux,b.schall2,b.weeks,b.hohler,b.coussot}. Usually, the existence of a yield stress is often associated to a flow behaviour governed by the Herschel-Bulkley equation relating the stress $\sigma$ to the strain rate~$\dot{\gamma}$: $\sigma=\sigma_0+C\dot{\gamma}^\beta$, where $\sigma_0$ is the quasi-static flow stress and $C$ and $\beta$ are constants~\cite{b.barnes}. The physical origin of this non-linear law is still under debate. On the one hand, it has been argued that these common rheological properties might reflect the presence of a glassy dynamics in these materials, which has been described by mean-field-like rheological models~\cite{b.sollich,b.falk,b.falk2,b.berthier3} inspired by a thermodynamical picture. Although these models can predict rather well the macroscopic properties of glassy materials, they are not able to take into account the spatial heterogeneities present in these systems, and thus to link the local heterogeneous dynamics to the macroscopic response. On the other hand, mesoscopic models propose to identify a small number of relevant parameters at a local level, and compute the macroscopic properties within cellular automaton simulations~\cite{b.martens,b.vandembroucq}, or integral equations~\cite{b.dahmen}. However, the local parameters used to extrapolate the rheological properties are characteristic times whose physical origin still remains unclear. This is why one has to investigate in details the microscopic, dissipative rearrangements in the system, and connect them to the global response of the material. Our hypothesis is that the characteristic times, if any, can be grasped by a geometrical description of the dynamics, and that characteristic lengthscales must be identified first.

In this paper we analyze the connection between the macroscopic rheology and the local microscopic dynamics by performing Molecular Dynamics simulations (MD) of a model disordered system submitted to a steady shear.

\section{Numerical Simulations}
\label{Tech}
We have investigated the rheological properties of a model amorphous silicon (a-Si) system consisting of $N_{at}=32768$ atoms contained in a cubic box with lengths $L_x= L_y = L_z$ of approximately 87 $\AA$. In order to study the local dynamics of the system, the sample has been sheared at constant shear rate $\dot{\gamma}$, ranging from $\dot{\gamma}=10^8$ to $\dot{\gamma}=10^{10}$ $s^{-1}$, by performing extensive Molecular Dynamics simulations at very low temperature ($T=10^{-5}K$) and imposed pressure P=2GPa, using the open source LAMMPS package~\cite{b.lammps}. The prescribed temperature corresponds to an athermal regime. It has been chosen in order to prevent thermal activation of energy barrier escapes, and to be sure that instabilities are driven only by the externally applied mechanical shear~\cite{b.rodney}. This athermal regime is particularly relevant when studying glasses far below the glass transition temperature, or in an attempt to transpose the results to amorphous assemblies of macroscopic particles (with micrometer size)~\cite{b.mobius,b.schall1,b.rodney}. The Si-Si interaction in the system studied here is described by the Stillinger-Weber potential~\cite{b.SW}, where we have tuned the prefactor of the three-body term  $\lambda$ to quantify the effect of local order, as we have done in our previous work~\cite{b.fusco}. The Stillinger-Weber potential is an emprical potential including two-body and three-body interactions, such that the total energy of the system is written as 
\begin{eqnarray}
E_{total}&=&\sum_{i<j}f(r_{ij}) + \lambda.\sum_{i<j<k} g(r_{ij},r_{ik},\theta_{jik})\\&+&g(r_{ji},r_{jk},\theta_{ijk})+g(r_{ki},r_{kj},\theta_{ikj})
\end{eqnarray}
with
\begin{eqnarray}
g(r_{ij},r_{ik},\theta_{jik})&=&\left(\cos\theta_{jik}+1/3\right)^2\\
&\times&\exp{\left(\alpha (r_{ij}-a)^{-1}+\alpha (r_{ik}-a)^{-1}\right)}
\end{eqnarray}
with $\alpha=1.20\AA$, $a=1.80\AA$.
The parameter $\lambda$ thus accounts for the bond's directionality: $\lambda=0$ corresponds to simple two-body interactions, while high $\lambda$ favors the local tetragonal order in our model materials ($\lambda=21$ is the original value proposed by Stillinger et al~\cite{b.SW} as an empirical model for a-Si).  Here we extend our results obtained on the quasistatically sheared a-Si sample to finite shear rates, with the aim to characterize the effects of the local dynamics on the plastic response of the system. The technical details of the preparation of the a-Si model have already been presented in Ref.~\cite{b.fusco}.

In our simulations the athermal limit and the constant pressure are ensured by applying respectively a thermostat and a barostat of Nose-Hoover type. The Molecular Dynamics simulations are performed with the time-reversible measure-preserving Verlet integrator and with a time step $\delta t=1fs$. The typical characteristic relaxation times used for the thermostat and the barostat, $\tau_{th}$ and $\tau_{bar}$ respectively, are $\tau_{th}=0.15\tau_{SW}$ and $\tau_{bar}=10\tau_{SW}$, where $\tau_{SW}$ is a characteristic Stillinger-Weber time defined as $\tau_{SW}=\sqrt{m.a^2/\epsilon}\approx 70 fs$, $m$ being the Si atom mass, and $a$ and $\epsilon$ characteristic length and energy scales appearing in the Stillinger-Weber potential. Note that this characteristic time is system dependent, such that larger particle sizes would give rise to larger time scales (and correspondingly smaller values for the shear rates). In the dynamical simulation, after each shear step, we perform a NVT equilibration run in order to equilibrate the temperature, followed by a NPT run which fixes the pressure to the desired value. The thermostatting and barostatting equilibration times are $1$ ps. After each $ps$ equilibration run, the shear is imposed homogeneously on the simulation box. The imposed shear step depends on the chosen shear rate: it will be $\delta\gamma=10^{-4}$ for the shear rate $\dot\gamma=10^8 s^{-1}$, $\delta\gamma=10^{-3}$ for $\dot\gamma=10^9 s^{-1}$ and so on. Under these conditions our system experiences an overdamped dynamics. We have also tested a different relaxation time for the thermostat, namely $\tau_{th}=1.5 \tau_{SW}$, checking that our results do not change significantly, provided that we equilibrate our system for a longer time. Moreover, each set of parameters has been applied to two different initial configurations of a-Si.

Finally, quasi-static simulations have also been performed at a constant pressure P=2GPa by relaxing the pressure through homogeneous compression and energy minimization, for each shear step. In the quasi-static case, the shear strain step is $\delta\gamma=10^{-3}$.

\section{Rheological response} 
\label{Rheo}
\begin{figure}
\resizebox{0.4\textwidth}{!}{\includegraphics{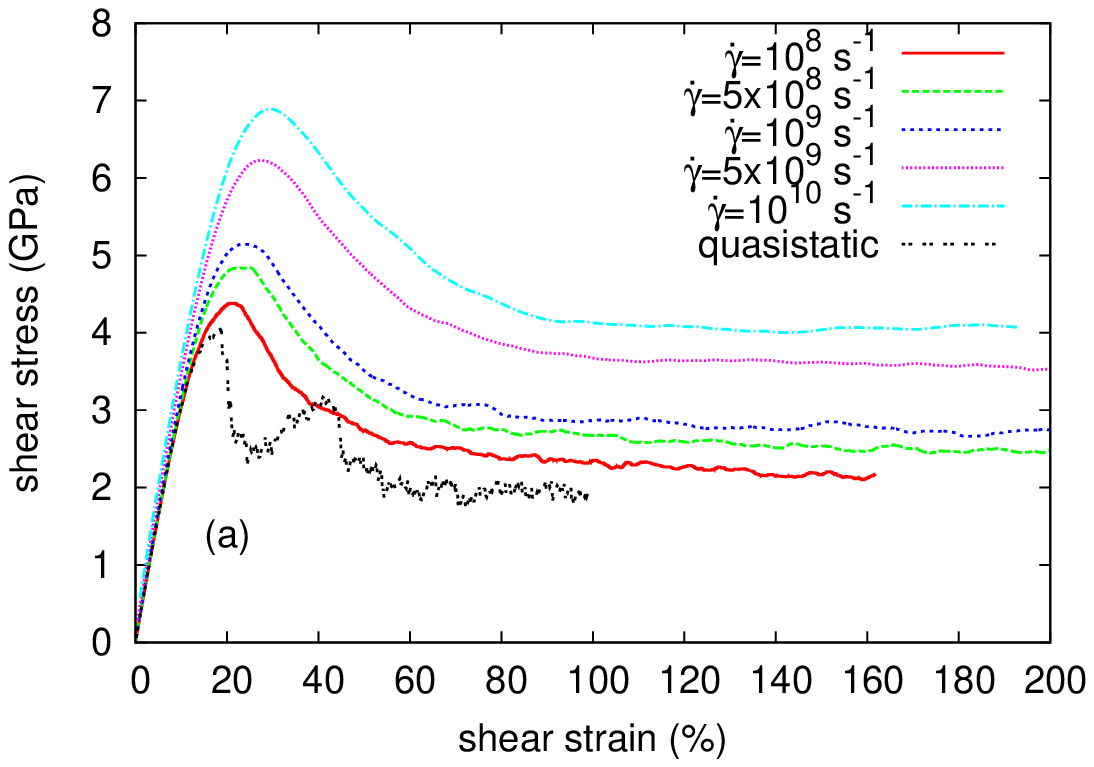}  }
\resizebox{0.4\textwidth}{!}{ \includegraphics{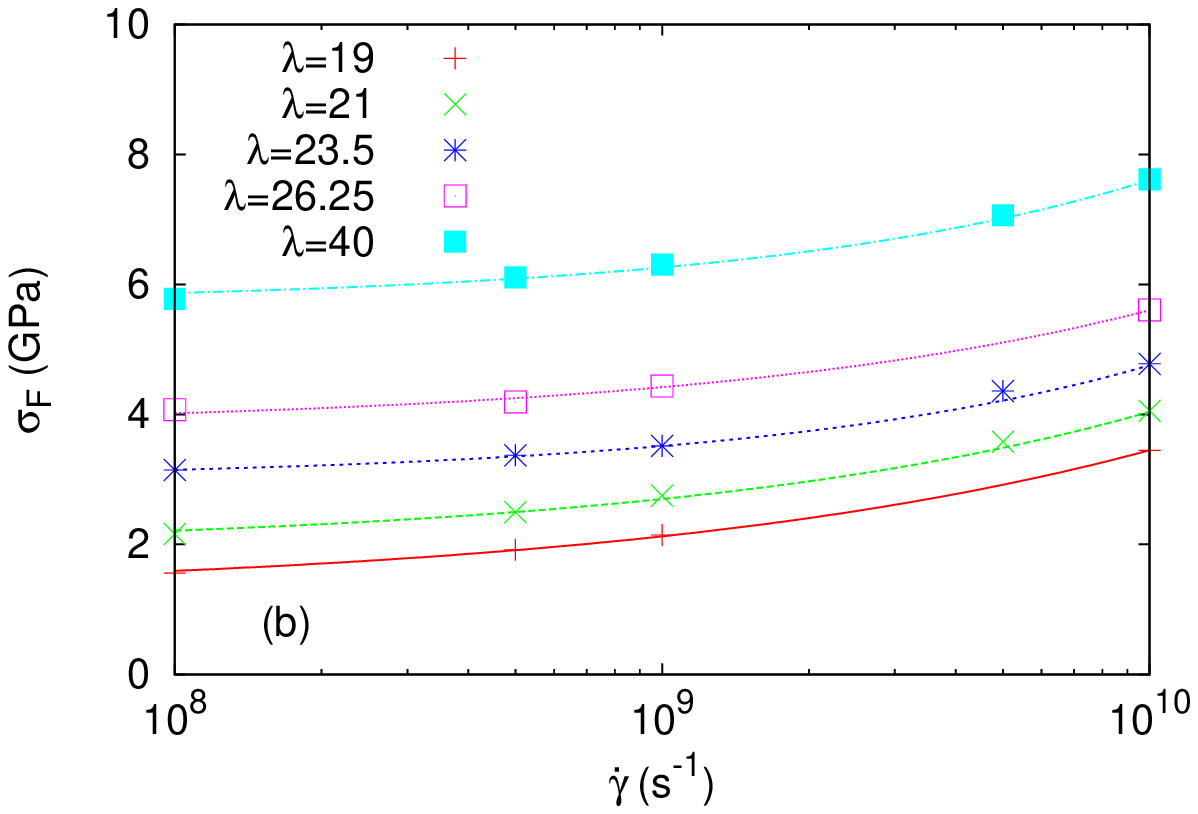} }
\resizebox{0.4\textwidth}{!}{ \includegraphics{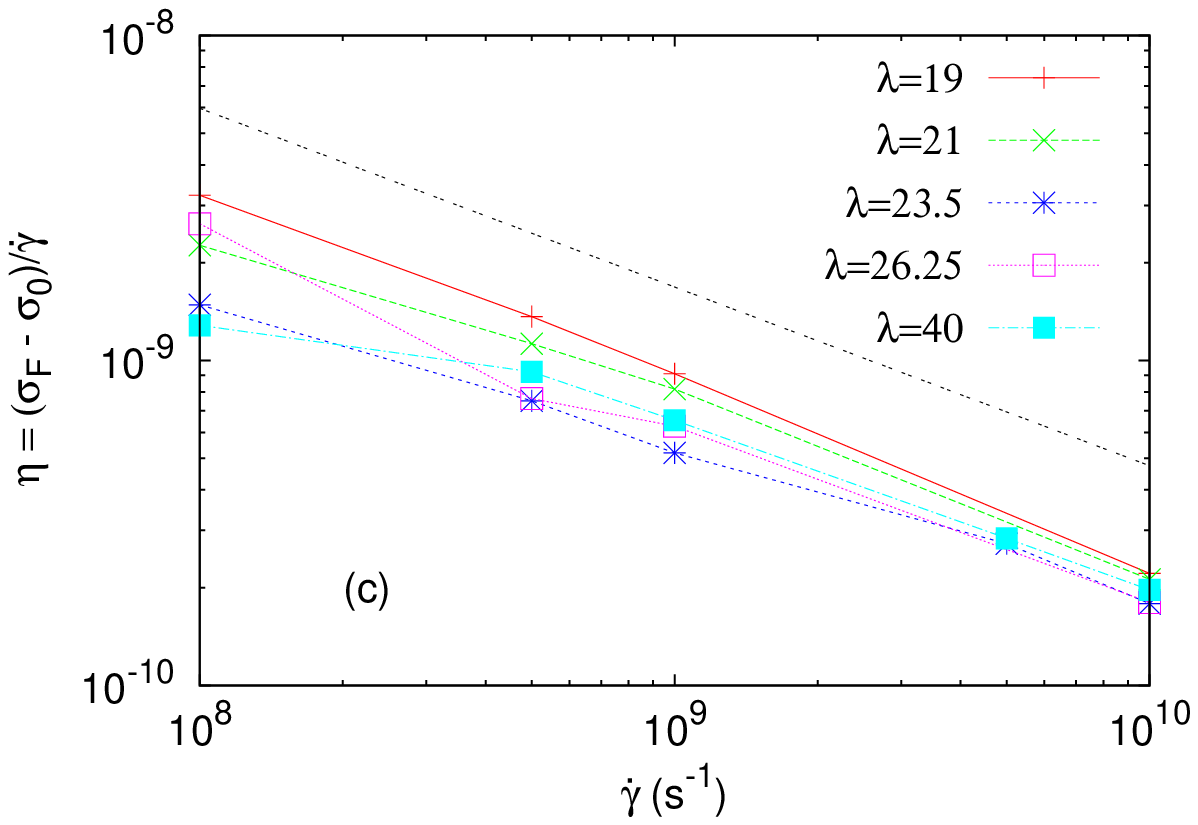} }
\resizebox{0.4\textwidth}{!}{ \includegraphics{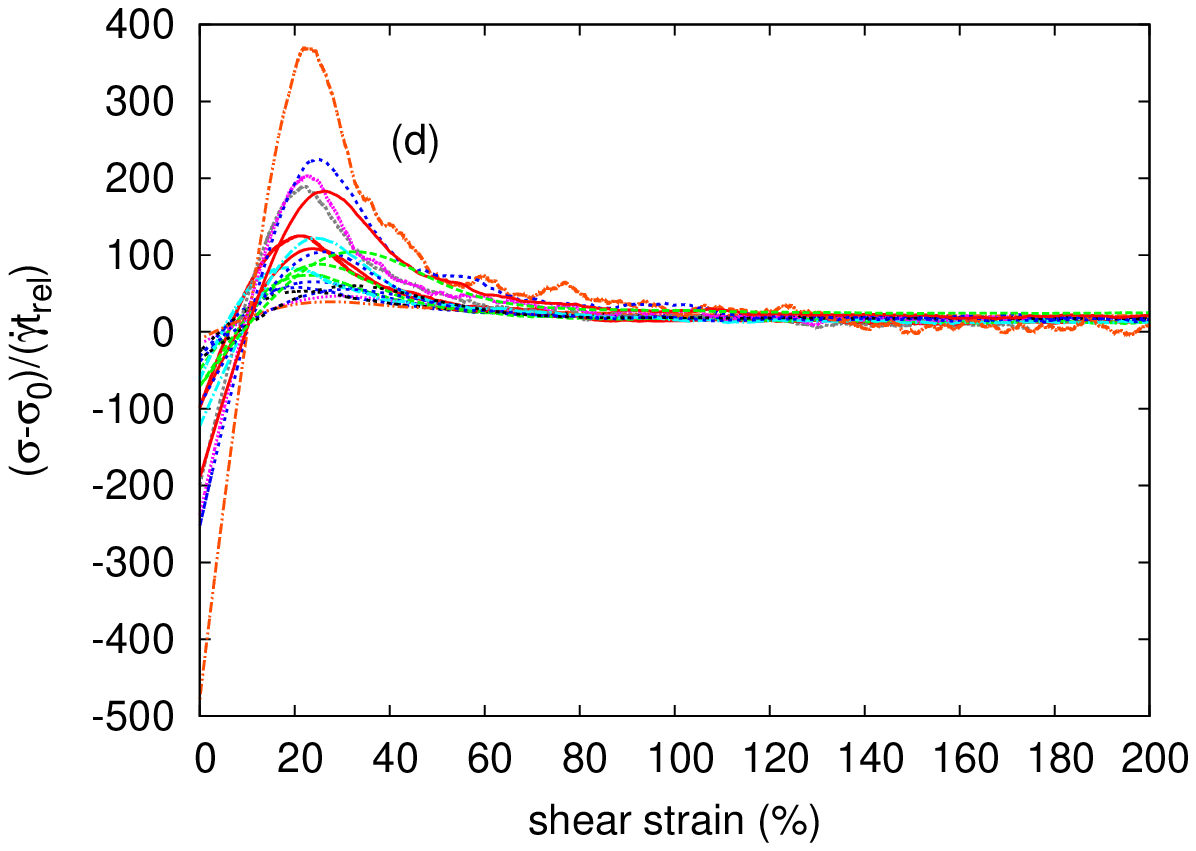} }
\caption{(a) Shear stress vs. shear strain for different shear rates and $\lambda=21$; for comparison the quasistatic case is also plotted. (b) Flow stress $\sigma_F$, defined as the average of the shear stress over the plastic plateau, as a function of the shear rate for different values of $\lambda$ (from bottom to top: $\lambda=19,21,23.5,26.25,40$). The solid lines are power-law fits to the points. (c) Viscous stress $\eta$ as a function of the shear rate for $\lambda=19,21,23.5,26.25,40$. (d) Rescaling of the shear stress as $(\sigma-\sigma_0)/\dot{\gamma}t_{rel}$ as a function of the shear strain for all the values of $\lambda$ ($\lambda=19,21,23.5,26.25,40$) and different shear rates ($\dot{\gamma}=10^8,5\cdot 10^8,10^9,5\cdot 10^9,10^10$ s$^{-1}$).}
\label{f.stress}
\end{figure}

\begin{figure}
\resizebox{0.6\textwidth}{!}{
\includegraphics{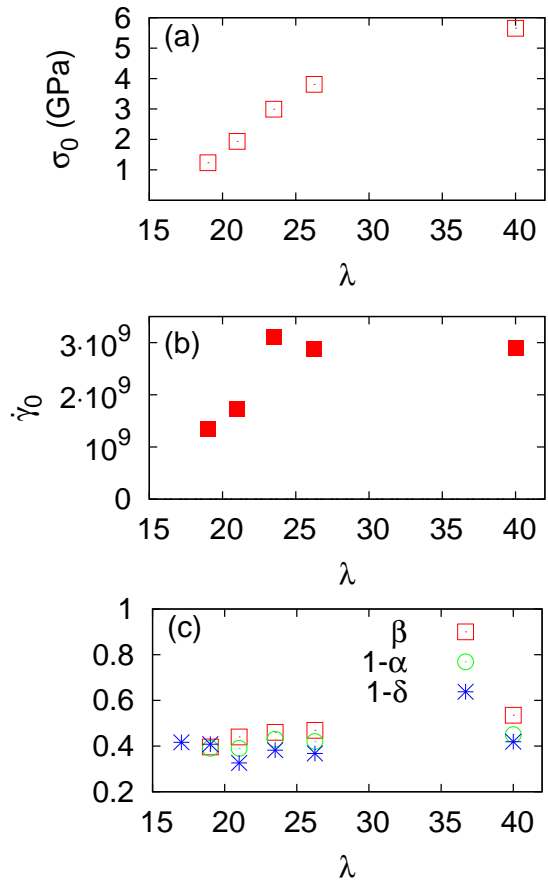}
}
\caption{(a) Quasi-static flow stress $\sigma_0$ as a function of $\lambda$. (b) Characteristic shear rate $\dot{\gamma}_0$ as a function of $\lambda$. (c) Values of the exponents $\beta$ (according to the Herschel-Bulkley relation), $1-\alpha$ (according to the power-law scaling of the relaxation time $t_{rel}$) and $1-\delta$ (according to the power-law scaling of the diffusion constant $D_T$) as a function of $\lambda$.}
\label{f.exp}
\end{figure}

The global mechanical response of the a-Si sample is probed by computing the shear stress as a function of the shear strain for different shear rates. This is illustrated in Fig.~\ref{f.stress} for different values of the shear rate and for a fixed value of $\lambda$ ($\lambda=21$).
In order to obtain a convergence of the shear stress (plastic plateau), we have deformed the system up to 200$\%$. From this figure we observe the characteristic mechanical behaviour of glassy materials: a linear part at low strain, a yield point, and a decrease of the stress up to a plateau corresponding to the plastic flowing regime. For comparison we have also shown the stress-strain relationship of the same system in the quasistatic limit. As already stressed in Ref.~\cite{b.tsamados}, we can see that the quasistatic stress-strain curve is the limiting case of the finite shear rates curves, i. e. the response converges to the quasistatic limit as the shear rate is progressively reduced. It is also obvious that the curves at finite shear rates are smoother than the one corresponding to the quasistatic procedure, since in the latter case the effects due to plastic collective rearrangements of atoms are more pronounced, causing the intermittent behaviour of the response and the appearance of ``jumps'' in the stress-strain curve, as thoroughly explained in Ref.~\cite{b.fusco}. The finite shear rate smears out these effects making the curves smoother and smoother as the shear rate increases.
In order to quantify the effect of the shear rate on the flow behaviour of the system we plot the flow stress $\sigma_F$, defined as the average of the macroscopic stress in the (last $20\%$ of the) plastic plateau, as a function of the shear rate $\dot\gamma$ in Fig.~\ref{f.stress}-b for different values of $\lambda$. A typical nonlinear flow curve following a Herschel-Bulkley behaviour is observed:
\begin{equation}
\label{e.herschel}
\sigma_F=\sigma_0+\left(\frac{\dot{\gamma}}{\dot{\gamma}_0}\right)^{\beta}.
\end{equation}
The  lines on figure \ref{f.stress}(b) are obtained from Eq.~(\ref{e.herschel}) where $\beta(\lambda)$ 
 and $\dot{\gamma}_0(\lambda)$ have been chosen as adaptable parameters.  $\sigma_0$ is taken from the quasistatic 
 calculations and increases as a function of $\lambda$ in agreement with our previous findings \cite{b.fusco}. 
The resulting characteristic shear rate $\dot\gamma_0$ in Eq.~(\ref{e.herschel}) increases with $\lambda$ and 
saturates at a finite value $\dot\gamma_0\approx 3.10^9 s^{-1}$ (see Fig.~\ref{f.exp}-b). The precise value of 
$\dot\gamma_0$ depends on the precision obtained on the fit of $\beta$ in a log-log plot. More important is the dependence on $\lambda$ of the measured exponent $\beta$ which is indeed very weak, as shown in Fig.~\ref{f.exp}-c, with values ranging from $0.4$ to $0.53$. Not surprinsingly, we could check that a fit with a fixed exponent $\beta=0.5$ also represents reasonably well our data. The values of the exponents $\beta$ for our system are similar to those found in colloidal glasses~\cite{b.besseling} and yield stress fluids~\cite{b.divoux}, and also metallic glasses~\cite{b.pelletier}. This universality of the exponent $\beta$ suggests a dynamical origin, independent on the specificities of the interatomic interactions.  
       
The values of the constants $\sigma_0$ and $\dot{\gamma}_0$ depend clearly on $\lambda$. In particular $\sigma_0$ (Fig.~\ref{f.exp}-a) increases as a function of $\lambda$ in agreement with the findings for the quasistatic case~\cite{b.fusco}. The characteristic shear rate $\dot\gamma_0$ in Eq.~(\ref{e.herschel}) increases with $\lambda$ and saturates at a finite value $\dot\gamma_0\approx 3.10^9 s^{-1}$ (see Fig.~\ref{f.exp}-b). The precise value of $\dot\gamma_0$ depends on the precision obtained on the fit of $\beta$ in a log-log plot. The dependence on $\lambda$ of the measured exponent $\beta$ is indeed very weak, as shown in Fig.~\ref{f.exp}-b, with values ranging from $0.4$ to $0.53$. The values of the exponents $\beta$ for our system are similar to those found in colloidal glasses~\cite{b.besseling} and yield stress fluids~\cite{b.divoux}, and also metallic glasses~\cite{b.pelletier}. This universality of the exponent $\beta$ suggests a dynamical origin, independent on the specificities of the interatomic interactions.

\section{Local Dynamics} 
\label{LocDyn}
\begin{figure}
\begin{center}
\resizebox{0.5\textwidth}{!}{
\includegraphics{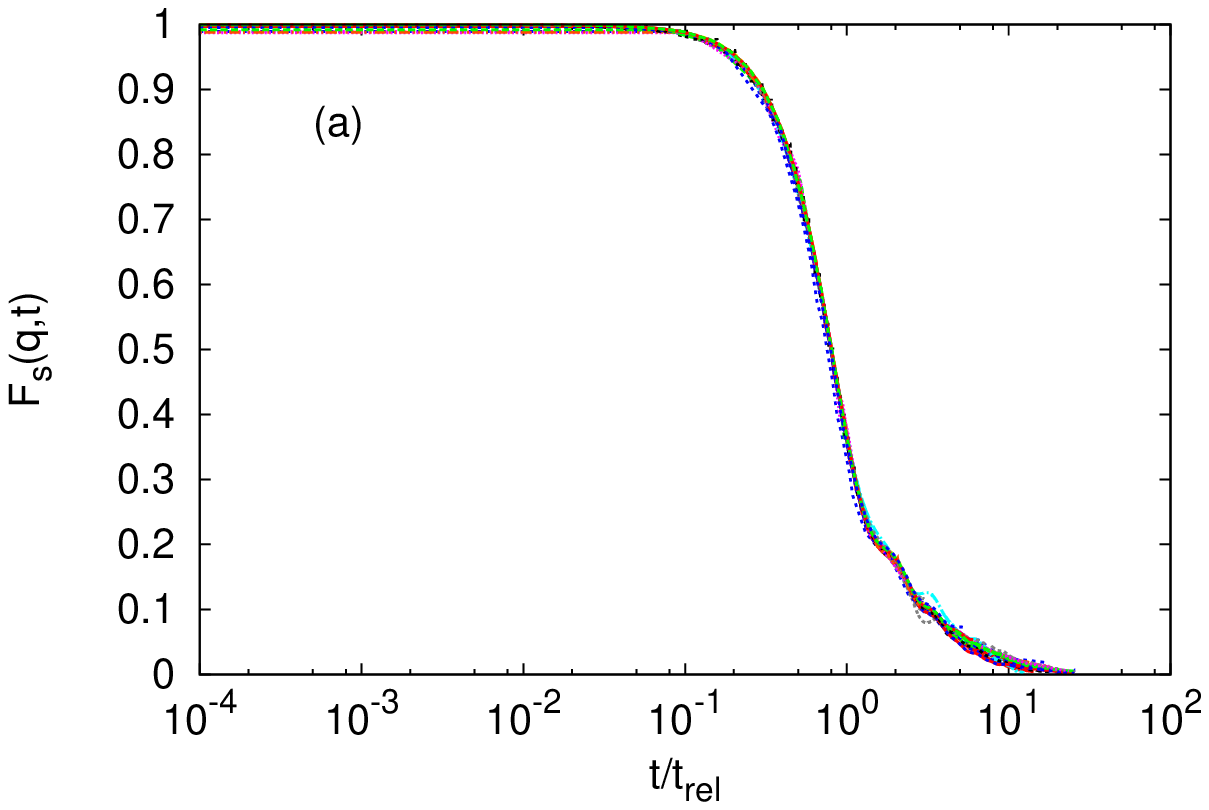}
}
\resizebox{0.5\textwidth}{!}{
\includegraphics{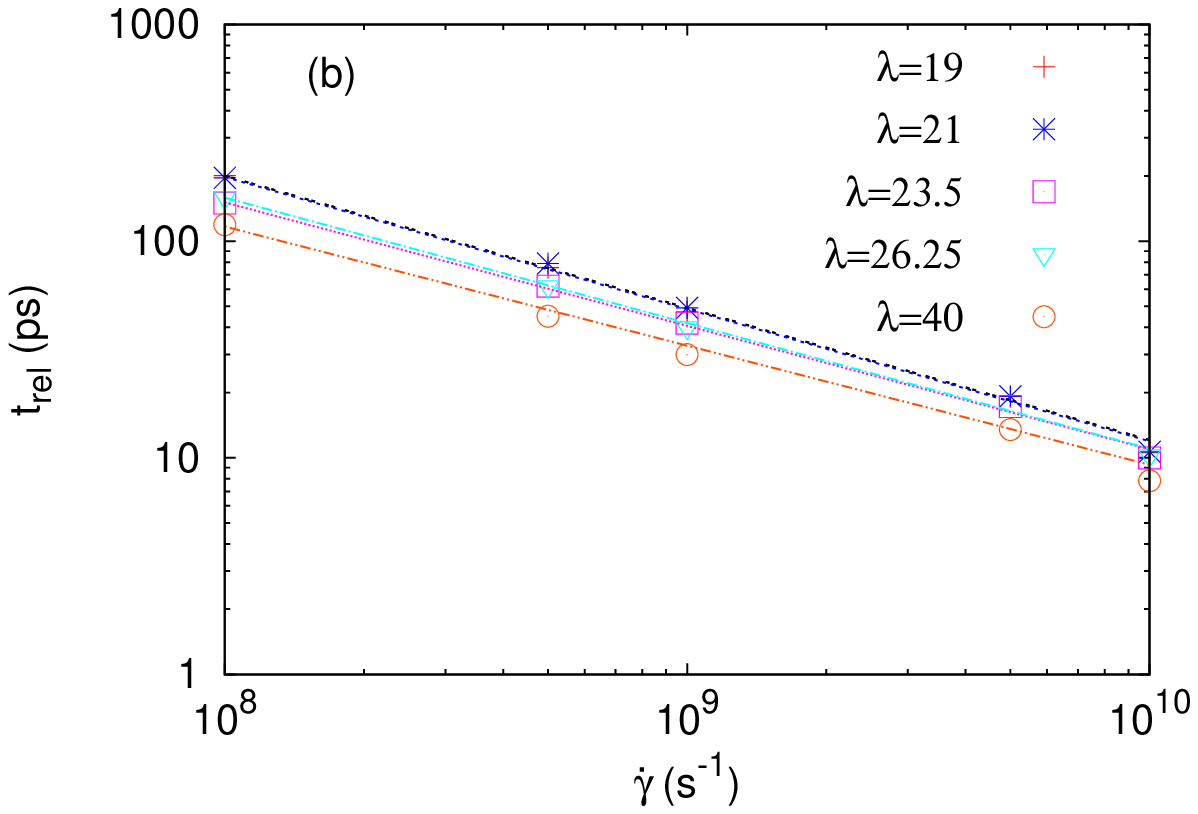}
}
\resizebox{0.5\textwidth}{!}{
\includegraphics{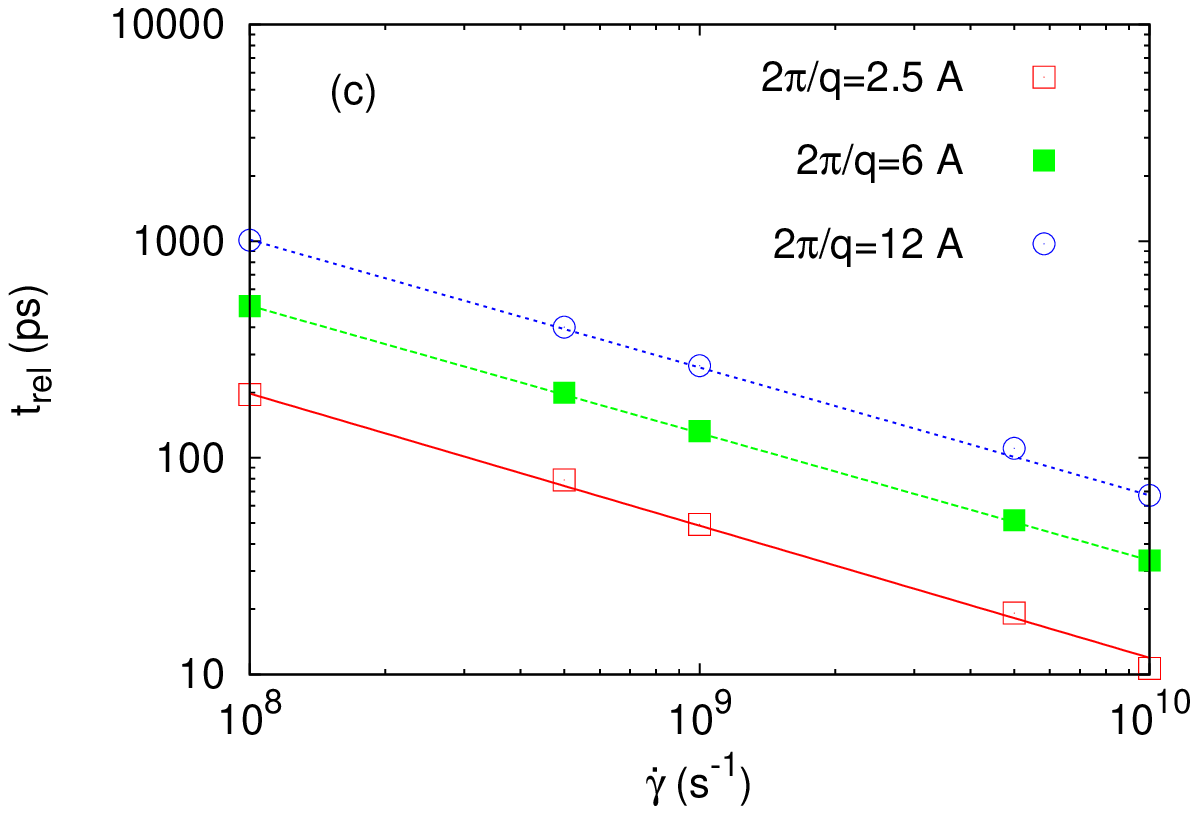}
}
\end{center}
\caption{(a) Rescaled self-intermediate scattering function for all the values of $\lambda$ ($\lambda=19,21,23.5,26.25,40$) and different shear rates ($\dot{\gamma}=10^8,5\cdot 10^8,10^9,5\cdot 10^9,10^10$ s$^{-1}$) and for three values of $q$ ($2\pi/q=2.5,6,12 \AA$): the curves superimpose very well on a master curve when time is rescaled by the relaxation time $t_{rel}$.
 (b) Relaxation time $t_{rel}$ as a function of the shear rate for different values of $\lambda$ at $2\pi/q=2.5 \AA$. The lines are power-law fit to the points. (c) Relaxation time $t_{rel}$ as a function of the shear rate for $\lambda=21$ and for three values of $q$ ($2\pi/q=2.5,6,12 \AA$). The lines are power-law fit to the points.}
\label{f.trel}
\end{figure}

In order to study the dynamics of the local density and to quantify the characteristic relaxation times, 
we study the self-intermediate scattering function (SISF) $F_s({\bf q},t)$:
\begin{equation}
F_s({\bf q},t)=\frac{1}{N_{at}}\sum_{i=1}^{N_{at}}\exp\{i{\bf q}\cdot[{\bf r}_{na,i}(t)-{\bf r}_{na,i}(0)]\}
\end{equation}
where ${\bf q}$ is the wave vector and ${\bf r}_{na,i}(t)$ is the non-affine displacement of atom $i$ at time $t$, 
obtained after substracting the affine displacement corresponding to a homogeneous shear strain
\begin{equation} 
{\bf r}_{na,i}(t)\equiv({\overline{\overline{1}}}-{\overline{\overline{\epsilon}}}).{\bf r}_i(t)
\end{equation}
where ${\overline{\overline{\epsilon}}}$ is the homogeneous shear strain applied to the simulation box, 
and ${\bf r}_i(t)$ is the position of atom $i$. 
The calculation of the SISF on the non-affine displacement 
field emphasizes the role of local strain heterogeneities on the local dynamics. For an isotropic system 
the SISF only depends on the modulus of $q$, thus $F_s=F_s(q,t)$.
We considered different values of $q$ : one is taken at the maximum of the structure factor $S(q)$ as suggested in \cite{b.berthier1}, this value of $q\simeq 2.5$ $\AA^{-1}$ does not considerably depend on $\lambda$  
in our systems and is associated to the first neighbour shell distance of $2\pi/q \simeq 2.5 \AA$; 
the second and the third values, $2\pi/q=6\AA$ and $2\pi/q=12\AA$ that span the typical extension of the average size of a plastic rearrangements. 
We see in Fig ~\ref{f.trel}-a that  $F_s(q,t)$ satisfies the time-shear superposition principle~\cite{b.berthier1,b.berthier2}, e. g. 
if we rescale the time by a quantity $t_{rel}(\dot\gamma,\lambda)$ that depends on the shear rate and on the parameter $\lambda$, 
all the SISF for different values of the shear rate and $\lambda$ collapse on a single master curve $F_s(q,t/t_{rel})$, for the three different $q$ chosen above. 
For a given $q$ value we represent 
 in Fig.~\ref{f.trel}-b $t_{rel}(\dot{\gamma},\lambda)$ in a log-log scale as a function of $\dot{\gamma}$. 
These results are consistent with a non-linear power law 
 dependence of $t_{rel}$ respect to $\dot{\gamma}$ :       
\begin{equation}
t_{rel}(\dot\gamma,\lambda)\sim\dot{\gamma}^{-\alpha}
\label{e-trelg}
\end{equation}
where the exponent $\alpha$ assumes values between 0.55 and 0.61, slightly depending on $\lambda$, and with a prefactor depending on $q$ (Fig.~\ref{f.trel}-c).

From Fig.~\ref{f.exp}-c it can be seen that the relation $\beta=1-\alpha$ is rather well satisfied, which is compatible with the hypothesis that the relaxation time is proportional to the viscosity $\eta$ defined from the viscous stress as
\begin{equation}
\label{e.eta}
\eta=(\sigma_F-\sigma_0)/\dot{\gamma}\sim\dot{\gamma}^{\beta-1}
\end{equation}
as already found in other glassy systems~\cite{b.tsamados,b.mobius}. A confirmation of the previous hypothesis is 
 shown on Fig.~\ref{f.stress}-d by the collapse of the rescaled viscosities 
$\eta/t_{rel}$ for all shear rates. 
This means that the effective stress $\sigma_{eff}=G_0.t_{rel}\dot{\gamma}$ obtained by the local dynamics, where $G_0$ is the shear modulus, is in good agreement 
with the global mechanical response of the material, which is a nontrivial result, that was already questioned in several systems 
(see e. g. Ref.~\cite{b.besseling}). 
Note however, that the local dynamics as studied here in the flowing regime is unable to take into account the quasi-static value of the flow, or solid yield, $\sigma_0$, as well as the transitional behavior (for $\gamma<40\%$).
 The same result was already mentioned in the experimental work of Ref.~\cite{b.mobius} on two-dimensional foams: the proportionality between the relaxation time measured from the local dynamics, and the apparent viscosity assumes that the effective (viscous) stress is measured after substracting the quasi-static value from the flow stress. 

\begin{figure}
\begin{center}
\vspace{0.5cm}
\resizebox{0.5\textwidth}{!}{ \includegraphics{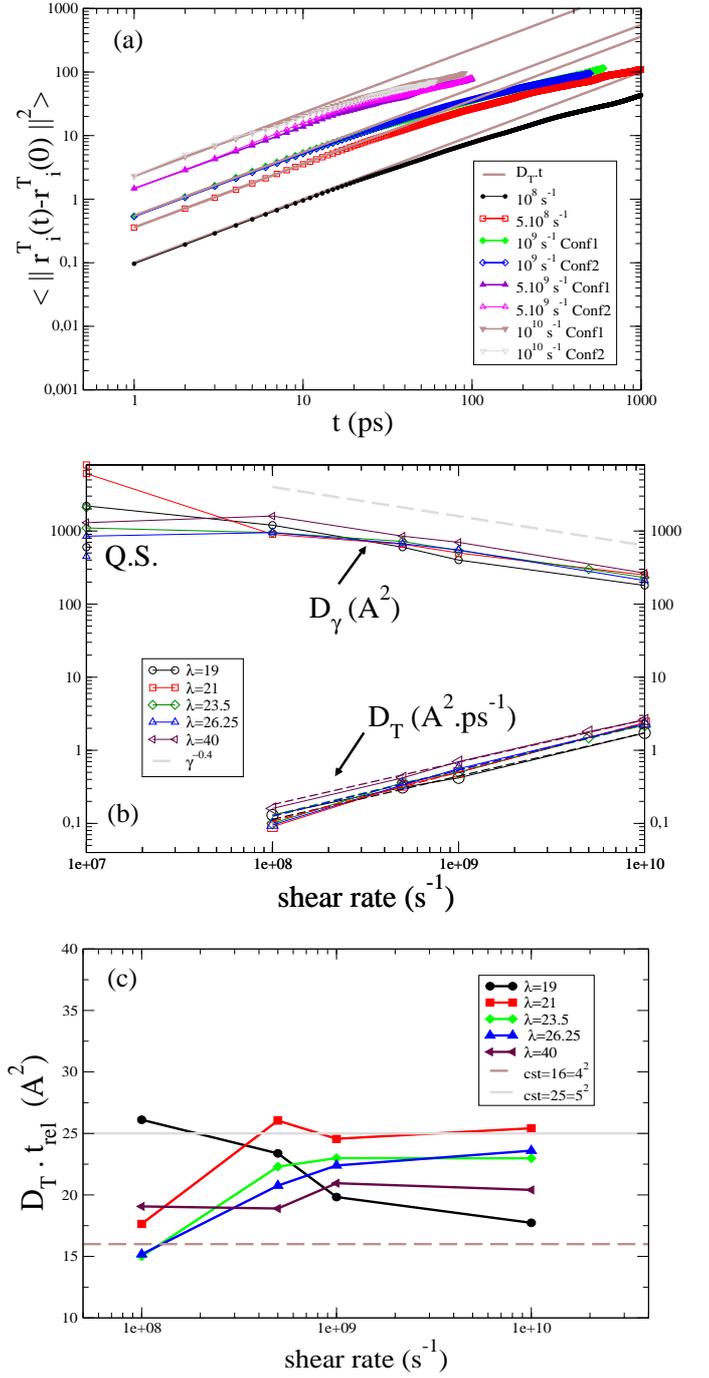} }
\end{center}
\caption{(a) Variance of the transverse motion as a function of time, for different shear rates and $\lambda=23.5$. (b) Dependence of the diffusion coefficients $D_\gamma$ and $D_T$ with the shear rate. The dashed lines correspond to the fit $D_T\propto\dot\gamma^\delta$ proposed in the text. (c) scaling $D_T .t_{rel}$ as a function of the shear rate for different values of $\lambda$.}
\label{f.diff}
\end{figure}

To check the above mentioned relation between apparent viscosity and local dynamics, we have also computed the apparent diffusion coefficient of atoms. Indeed, due to plastic deformation, and even in the athermal regime, atoms undergo diffusive motion~\cite{b.tanguy}. The variance of the transverse motion (perpendicular to the shear direction) is proportional to the time elapsed
\begin{equation}
<(r^T_i(t)-r^T_i(0))^2>=D_T\cdot t
\end{equation}
This relation holds very well in our 3D system, as long as the displacement is not too large, as can be seen on Fig.~\ref{f.diff}-a.
From the calculation of $D_T(\dot{\gamma},\lambda)$ displayed in Fig.~\ref{f.diff}-b we propose the following scaling law for 
 $D_T$ :
\begin{equation} 
D_T\propto\dot\gamma^\delta
\label{e.treld}
\end{equation} 
The Stokes-Einstein relation for the stochastic motion of particles would give $D_T\propto 1/\eta$, that is $\delta=1-\beta$. As it can be checked on Fig.~\ref{f.exp}-b,
this relation holds reasonably well for all the values of $\lambda$ studied, emphasizing the role of the non-affine dynamics on the dissipative behaviour of our material. 
The consistency of this last scaling law can be evaluated respect to the previously defined $\alpha$ and $\beta$ exponent by calculating 
$D_T.t_{rel}(\dot\gamma,\lambda)$ which should reduce to a constant value. This property is rather well verified 
considering the spread of the corresponding diffusion length $\sqrt{D_T.t_{rel}}$ which is of the order of 1 \AA, as shown in  Fig.~\ref{f.diff}-c.
Note that in the quasi-static regime studied in~\cite{b.fusco}, a finite diffusive coefficient $D_\gamma$ can be defined by relating the transverse motion to the applied strain (and no more to the time elapsed). The relation $D_\gamma\equiv D_T/\dot\gamma$ imposes a saturation of $D_T/\dot\gamma$ at small shear rates. In our present simulations, the value of $D_\gamma$ measured for the smallest shear rate is already very close to the quasi-static value (Fig.~\ref{f.diff}-b). This QS value acts as a upper cut-off, whose precise value can however be very system-dependent~\cite{b.tsamados}.

The analysis presented in this section allows to identify time scales that are relevant to describe the local atomistic dynamics, through the study of the temporal correlations in atomic positions, averaged over the whole system. Here a single time scale $t_{rel}$ emerges. It however does not allow to understand the physical origins of the corresponding relaxational processes.
In the following we argue that the non-linear shear rate dependence of the relaxation time could be explained by a simple model based on the competition between the nucleation and the diffusive propagation of plastic events.

\begin{figure}
\begin{center}
\resizebox{0.5\textwidth}{!}{
\includegraphics{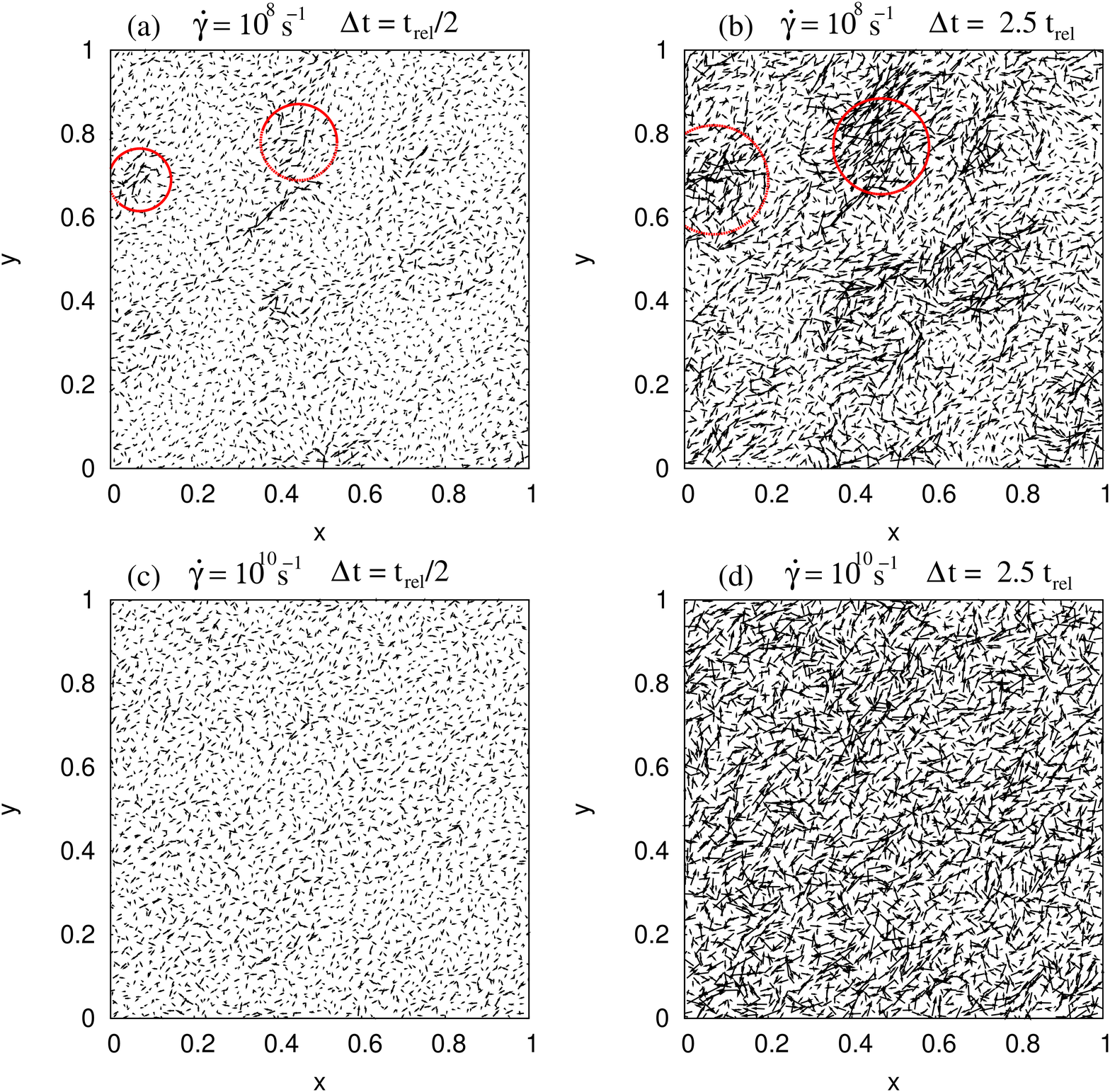}
}
\end{center}
\caption{Snapshots of the non-affine displacement field in the $(x,y)$ plane in a narrow $z$ region ($0.45<z<0.55$) for different times $\Delta t=\Delta\gamma/\dot\gamma=0.5 t_{rel}$ and $\Delta t=2. t_{rel}$, for $\dot{\gamma}=10^8$ s$^{-1}$ (top panel) and  $\dot{\gamma}=10^{10}$ s$^{-1}$ (bottom panel). $\lambda=19$. Coordinates are in relative units between 0 and 1. The arrows of the atomic non-affine displacements are magnified by a factor 5.}
\label{f.snapshots}
\end{figure}

\section{Analysis of dissipative events}
\label{Ana}

\begin{figure}
\begin{center}
\resizebox{0.4\textwidth}{!}{ \includegraphics{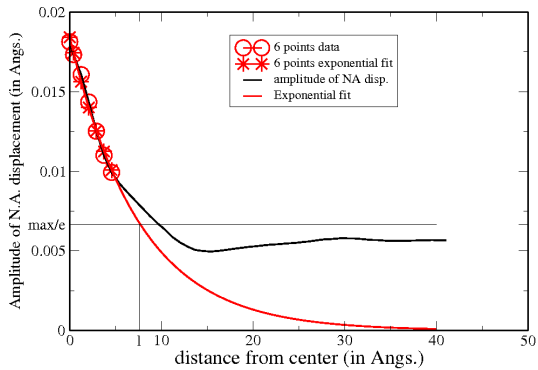} }
\end{center}
\caption{Angular average of the amplitude of a non-affine displacement field centered on its local maximum, and of the corresponding atomistic interaction energy as a function of the distance to its local maximum. Dashed line: exponential fit on the first 6 points of the curve. This is the fit used to determine the sizes $l(\Delta\gamma)$ discussed in the text for the different $\dot\gamma$ and $\lambda$ values.}
\label{f.fit}
\end{figure}

Fig.~\ref{f.snapshots} presents a snapshot of the non-affine displacement for two values of $\dot{\gamma}$ at the same deformation step and for two different fractions of the time interval $t_{rel}$. The non-affine displacement field is defined as the departure from the simple displacement due to the homogeneous strain. It is obtained during the constant pressure-constant temperature dynamics of the sample. It is possible to compute a non-affine displacement field for different strain (or time) intervals by comparing the final and the initial particle positions, and substracting the total contribution due to the accumulated homogeneous shear. We have shown that the local maxima in the amplitude of the non-affine displacement field coincide with the local maximum changes in the atomistic interaction energy. In the quasi-static regime, it corresponds also to the maximum dissipated energy during go-and-reverse simulation run~\cite{b.fusco}. We thus used the determination of the non-affine displacement fields as an indicator for visco-plastic events. Fig.~\ref{f.snapshots} shows that the non-affine displacements grow with time with a vortex-like structure. It is also clear from Fig.~\ref{f.snapshots} that for low $\dot{\gamma}$ the plastic events are fewer and more extended, while for higher $\dot{\gamma}$ there are many small plastic events scattered throughout the system.

\begin{figure}
\resizebox{0.4\textwidth}{!}{
\includegraphics{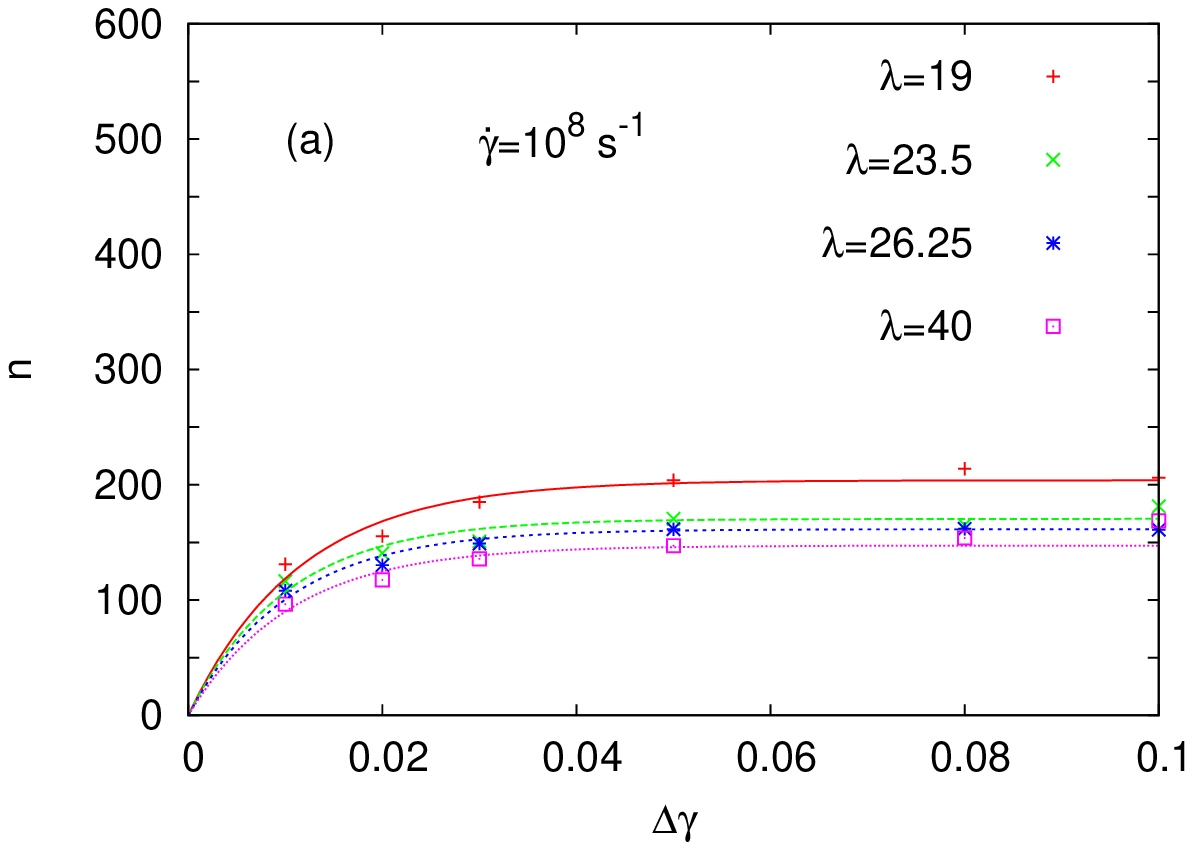}
}
\resizebox{0.4\textwidth}{!}{
\includegraphics{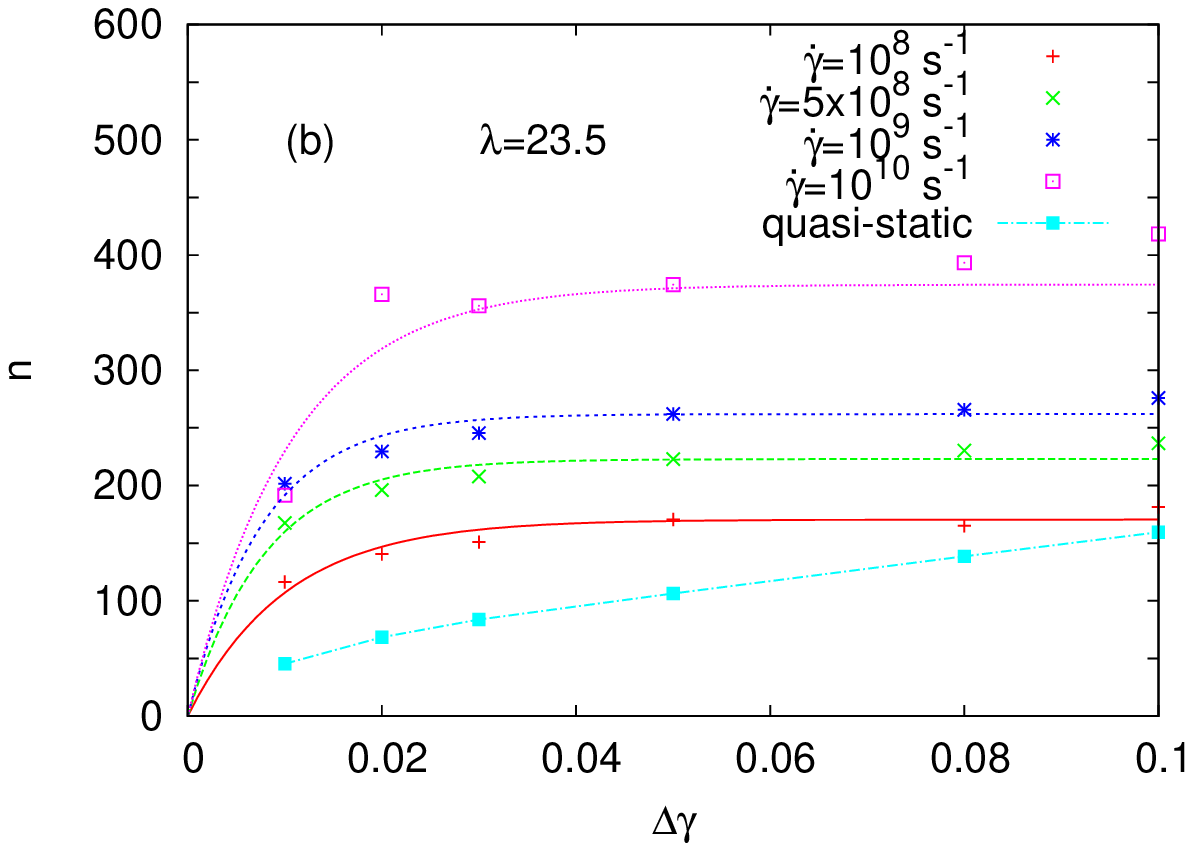}
}
\resizebox{0.4\textwidth}{!}{
\includegraphics{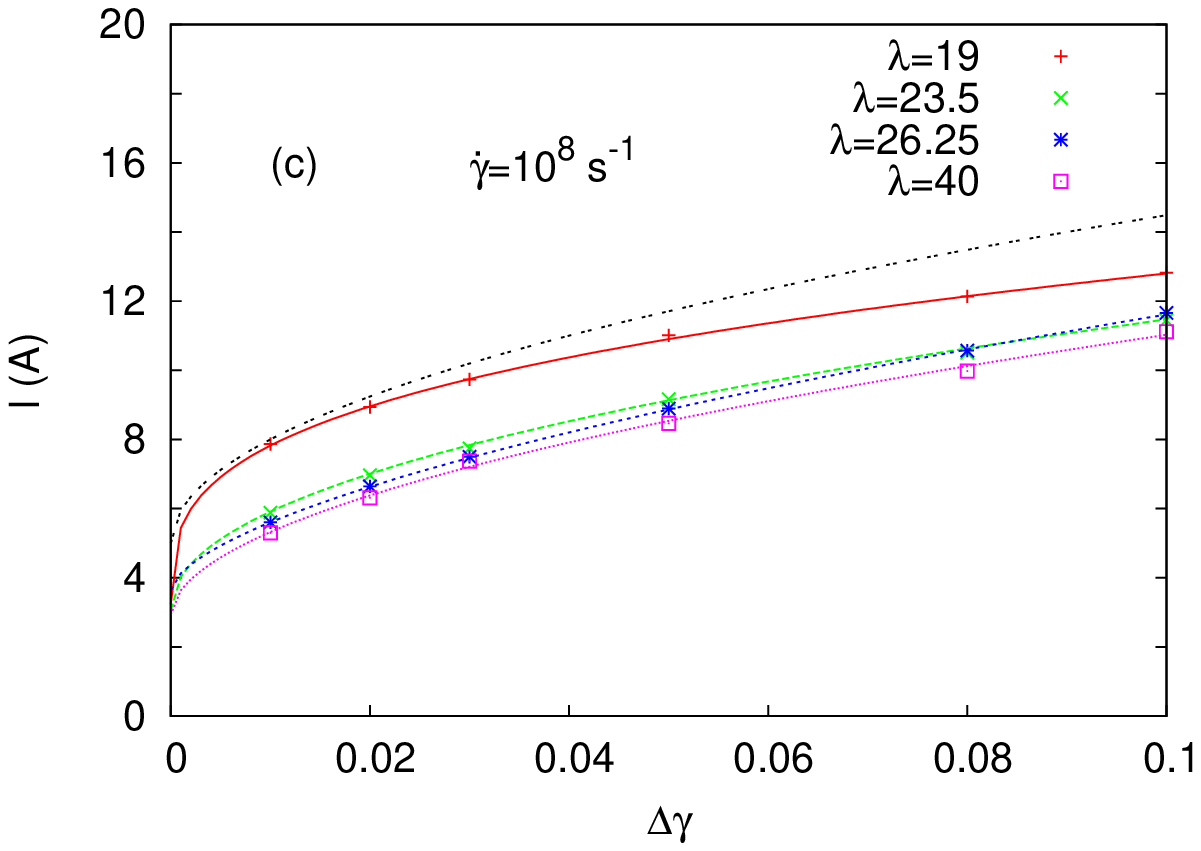}
}
\resizebox{0.4\textwidth}{!}{
\includegraphics{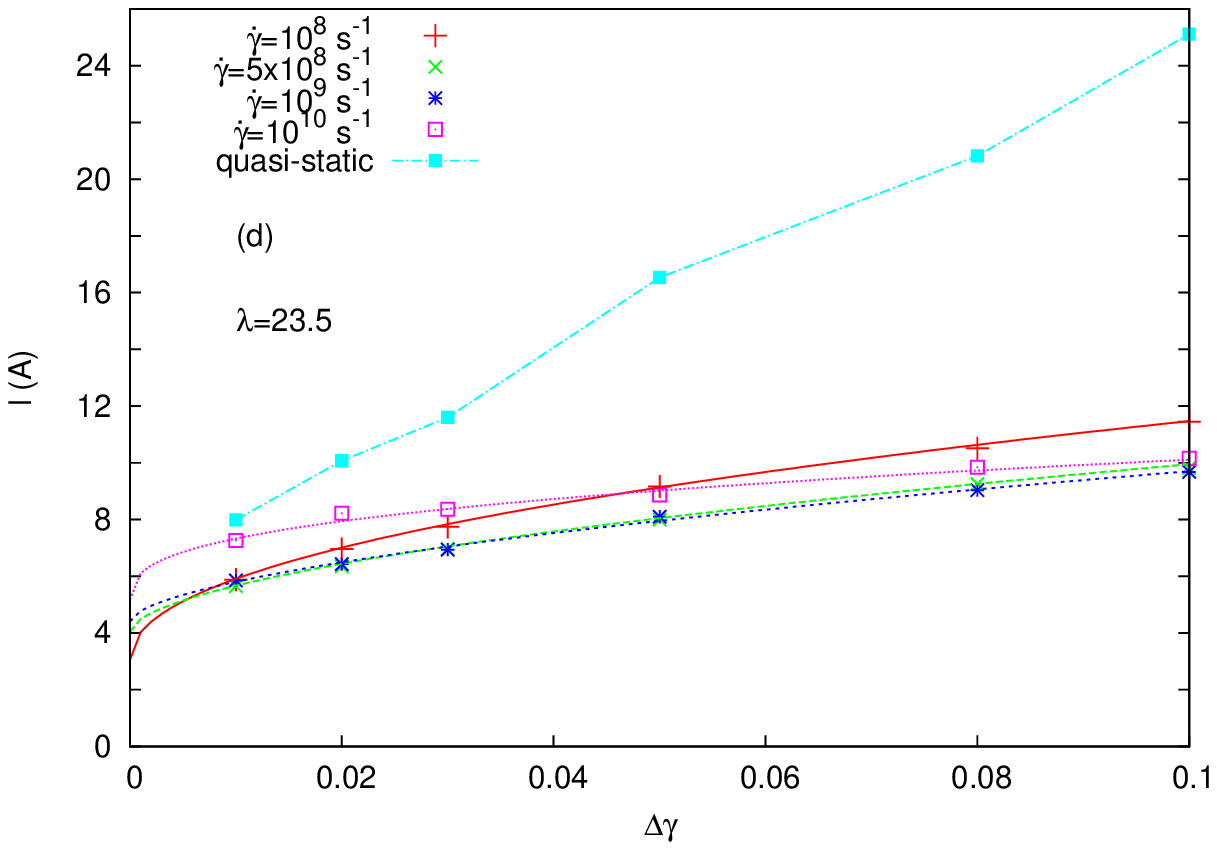}
}

\caption{Dependence of the average number of plastic events (a)-(b) and of the average size of plastic events (c)-(d) on $\Delta\gamma$ for different values of $\lambda$ at fixed $\dot{\gamma}=10^8$ s$^{-1}$ (a)-(c) and for different values of $\dot{\gamma}$ at fixed $\lambda=23.5$ (b)-(d). The black dashed lines without symbols in (c) is a reference curve of the form $l\propto \Delta\gamma^{0.5}$ to guide the eye. Lines correspond to the fits discussed in the text, despite for the quasi-static case that is not the purpose of this article.}
\label{f.nattract_avsize}
\end{figure}

We have analyzed the plastic events as a function of the shear strain by identifying the maxima in the amplitude of the non-affine displacement field. The maxima have been determined as the attractors of this field by using the same procedure outlined in Ref.~\cite{b.fusco}.
From this analysis we evaluated the number of plastic rearrangements (n) in a given interval $\Delta{\gamma}$ and the average size (l) of the rearrangements as follows. To focus on the most relevant events, we first calculated for all the attractors an integrated non-affine amplitude over the whole basin associated to each attractor. Then we retained in our analysis all the rearrangements whose integrated non-affine amplitudes were greater than 20\% of the maximum value.
The size of the rearrangements has then been determined using an exponential fit of the angular average of the local non-affine displacement field, restricted to distances $r$ very close to the maximum ($r<6\AA$). Example of such a fit is shown in Fig.~\ref{f.fit}. 
The maximum range of $6\AA$ corresponds to the typical size of the core of the plastic rearrangements ~\cite{b.fusco}. 
It takes into account the plastic local reorganisation in the core of a rearrangement, while neglecting the long-range elastic decay
surrounding the plastic pinch.
 At large $\Delta\gamma$ a saturation can also appear at large distance, due to the increase of activity and finite size effects, as will be discussed later. This saturation appears beyond the restricted fit interval that allows us to isolate the visco-plastic centers.   
For each strain interval $\Delta\gamma$ between two configurations in the plastic plateau, we have determined the number of plastic events and their average size. The average number $n$ of plastic events as a function of the strain interval $\Delta\gamma$ between two successive configurations is shown in Fig.~\ref{f.nattract_avsize}-a for different values of $\lambda$ (for $\dot{\gamma}=10^8$ s$^{-1}$) and in Fig.~\ref{f.nattract_avsize}-b for different values of $\dot{\gamma}$ (for $\lambda=23.5$). The average is obtained on different strain origins in the plastic plateau, along the last $100\%$ of strain deformation. It appears that $n$ increases first linearly with $\Delta\gamma$, and tends to saturate at large $\Delta\gamma$ ($\Delta\gamma>\Delta\gamma_c$). The saturation is due to finite size effects: from a given density of events, the number of maxima does not change anymore; any additional event will fall in the attraction basin of another maximum, and contribute to increase the overall average displacements. From that figure, it is clearly shown that $n$ increases by decreasing $\lambda$ and/or by increasing the shear rate. In order to get a quantitative description of this measurement, we used an exponential fit to describe the saturation effect, corresponding to the linear increase at small $\Delta\gamma$ followed by the finite size saturation above a critical value $\Delta\gamma_c\approx 1\%$ depending only on the system size
\begin{equation}
n=n_0\cdot\left(1-\exp(-\Delta\gamma/\Delta\gamma_c)\right)
\label{eq.n}
\end{equation}
The fits are shown in Fig.~\ref{f.nattract_avsize} for different values of the shear rate and of $\lambda$. They are quite good, except for the highest value of the shear rate, where the number of distinct visco-plastic centers is very high already from the very beginning. The prefactor $n_0$ depends on $\lambda$ as well as on $\dot\gamma$. Its low shear rate limit should coincide with its quasi-static value $n_{QS}(\lambda)$. It is thus natural to look for a non-linear fit of the form
\begin{equation}
n_0=n_{QS}(\lambda)+n_1(\lambda)\cdot\dot\gamma^{n_e(\lambda)}
\end{equation}
with only two unknown parameters $n_1(\lambda)$ and $n_e(\lambda)$. The different parameters $n_0:w
$, $n_1$, $n_e$ and $\Delta\gamma_c$ are summarized in Table~\ref{table.coeffn}. It is shown that $n_e>0$ and increases slowly with $\lambda$, while $n_1$ decreases exponentially with $\lambda$. The positive sign of $n_1$ and of $n_e$ ensures the increase of $n$ with the shear rate $\dot\gamma$ as stated before.

\begin{table*}{}
\begin{center}
\begin{tabular}{|c|cccc|c|c|cccc|}
\hline\noalign{\smallskip}
                      & $n_0$& & & & $n_1$& $n_e$& & & $\Delta\gamma_c$ &   \\
\noalign{\smallskip}\hline\noalign{\smallskip}
          & $\dot\gamma=10^8s^{-1}$& $5.10^8s^{-1}$& $10^9s^{-1}$& $10^{10}s^{-1}$&
& & $\dot\gamma=10^8s^{-1}$& $5.10^8s^{-1}$& $10^9s^{-1}$& $10^{10}s^{-1}$\\
\noalign{\smallskip}\hline\noalign{\smallskip}

$\lambda=19$    & 203.9 &257.0 &265.4 &347.5 &4.44 &0.175 &0.011 & 0.011 & 0.008 & 0.010\\
$\lambda=23.5$  & 170.5 &222.9 &262.0 &374.3 &0.58 &0.27 &0.010 & 0.008 & 0.008 & 0.010\\
$\lambda=26.25$ & 161.5 &223.3 &252.3 &371.1 &0.42 &0.28 &0.010 & 0.009 & 0.0075 & 0.008\\
$\lambda=40$    & 147.2 &198.8 &229.2 &393.0 &0.06 &0.37 &0.010 & 0.009 & 0.0075 & 0.010\\
\noalign{\smallskip}\hline
\end{tabular}
\end{center}
\caption{Coefficients obtained in the fit of the average number $n$ of distinct plastic rearrangements as discussed in the text.}
\label{table.coeffn}
\end{table*}

\begin{table*}{}
\begin{center}
\begin{tabular}{|c|cccc|cccc|cccc|c|}
\hline\noalign{\smallskip}
                      & $l_0$& & & & $l_1$& & & & $n_l$& & & & $160.\dot\gamma^{-0.5/3}$    \\ 
\noalign{\smallskip}\hline\noalign{\smallskip}
$\dot\gamma$          & $\lambda=19$& $23.5$& $26.25$& $40$&
$\lambda=19$& $23.5$& $26.25$& $40$&$\lambda=19$& $23.5$& $26.25$& $40$&\\ 
\noalign{\smallskip}\hline\noalign{\smallskip}
$10^8s^{-1}$  & 6.9 &5.1 &4.35&3.9 &27.01 &35.38&38.65&32.65&0.503&0.615&0.61 &0.53 & 6.98\\
$5.10^8s^{-1}$& 5.28&3.9 &3.33&2.99&22.69 &22.55&20.49&22.7 &0.393&0.461&0.44 &0.465& 5.31\\
$10^9s^{-1}$  & 4.7 &3.47&3.01&2.76&21.77 &19.99&18.15&19.14&0.562&0.42 &0.38 &0.42 & 4.72\\
\it{$10^{10}s^{-1}$} & 3.4 &2.8 &2.02&1.9 &17.032&14.66&13.19&10.33&0.19 &0.188&0.183&0.165& 3.20\\
\noalign{\smallskip}\hline 
\end{tabular}
\end{center}
\caption{Coefficients obtained in the fit of the average size $l$ of plastic rearrangements as discussed in the text.}
\label{table.coeff}
\end{table*}

Figs.~\ref{f.nattract_avsize}-c-d display the average size $l$ of plastic events as a function of $\Delta\gamma$ for different values of $\lambda$ and $\dot{\gamma}$, along the last $100\%$ of strain deformation. The size $l$ is given by the small scale exponential fit of the non-affine displacement field centered on each attractor. The fitted size evolves with increasing strain intervals, as can be seen in Fig.~\ref{f.snapshots}. It is seen here that the size $l$ increases approximately diffusively with the strain interval. It means, that the accumulation of visco-plasticity in the same attraction basin contributes to a diffusive increase of its apparent size.
We can fit the average size of plastic events with a power-law of the type
\begin{equation}
l=l_0(\lambda,\dot\gamma)+l_1(\lambda,\dot\gamma)\Delta\gamma^{n_l}
\label{eq.l}
\end{equation}
with values given in Table~\ref{table.coeff}.
The exponent $n_l$ slightly depends on the values of $\lambda$ and $\dot{\gamma}$, but (despite for the highest shear rate) it can be considered that it is close to $0.5$, suggesting a diffusive behaviour of the growth of the visco-plastic centers, as stated before. Since the value of $l$ results from the accumulation of large non-affine displacements due to plastic activity around a given center, it can be considered as a measurement of some kind of avalanche size, or progressive unfolding of the displacements, as a function of $\Delta\gamma$ around an initial plastic event. On the other side, $n$ corresponds to the number of simultaneously generated attractors. These measurements confirm the existence of two different populations of plastic events: {\it avalanche-like} events that propagate closely from an initial center and that give rise to a diffusive increase of $l$, and {\it separated} events that give rise to the linear increase of $n$ as a function of $\Delta\gamma$.
The competition between these two kinds of events is responsible for the relaxation dynamics of the system.

\section{Simple model}
\label{SimMod}
In fact, we propose as in~\cite{b.tsamados} to determine the relaxation time scale $t_{rel}$ as the time at which the size of the plastic events reaches the distance between independently generated events. It corresponds to the time at which the system is entirely rejuvenated by plastic rearrangements. This time is given by equating $l$ and $L/n^{1/d}$ where $d$ is the dimension of space ($d=3$ for point-like defects, $d=2$ for dislocation-like defects), and $L$ is the system size.
Using Eq.(~\ref{eq.n}) and Eq.(~\ref{eq.l}) and replacing $\Delta\gamma$ by $\dot\gamma.t_{rel}$, the above relation between $n$ and $l$ gives to the first order in $\Delta\gamma$ in the low deformation limit:
\begin{equation}
t_{rel}=\frac{\Delta\gamma_c\cdot L^d}{\dot\gamma\cdot l_0(\lambda,\dot\gamma)^d\cdot n_0}
\end{equation}
Since $n_0$ increases with $\dot\gamma$, this expression shows clearly that in the absence of any shear rate dependence in $l_0$, it is impossible to recover the Herschel-Bulkley behavior, where $t_{rel}\propto\eta\propto\dot\gamma^{\beta-1}$ with $0<\beta$. The Herschel-Bulkley behaviour is recovered if $l_0(\lambda,\dot\gamma)\propto\dot\gamma^{-\beta/d}$ decreases with the shear rate, in agreement with our measurement (see Table~\ref{table.coeff}). Thus this shows that the Herschel-Bulkley exponent results from the avalanche dynamics of closely related events in our systems, as already suggested in~\cite{b.caroli}.
Moreover, as stated before, $t_{rel}$ describes the relaxational processes only in the viscous regime, that is above the quasi-static yield. Its measured proportionality to $\eta$, discussed before, would give $t_{rel}\propto (\sigma_F - \sigma_0)^{(\beta-1)/\beta}$ by combining Eq.~\ref{e.herschel} and Eq.~\ref{e.eta}. Our work suggests that this non-linear dependence would result from the non-linear decrease of the unfolding of visco-plastic events with the shear rate.

\section{Conclusion}
\label{Conclusion}
In this paper, we have shown evidence of the existence of two types of plastic rearrangements ({\it avalanche-like} events and {\it separated} events), and of the crucial role played by the competition between nucleation and propagation of the plastic activity in the rheological behaviour of overdamped systems characterized by different interatomic interactions. We focus more precisely on the role of bond directionality in the rheological behaviour or amorphous materials in the very low temperature regime where local dissipative processes allow an efficient heat extraction. In this regime, the non-linear rheological properties can be explained by a single relaxation time scale, that we relate to the non-linear shear rate dependence of the avalanche properties (unfolding) of visco-plastic rearrangements. This mechanisms gives a dynamical explanation for the exponent of the Herschel-Bulkley law, while the specificities of bond interactions would be included mainly in the yield process. This explanation suggests to consider not only the number of plastic rearrangements, but also the size evolution of plastic rearrangements in mesoscopic modelling of plasticity of amorphous materials of any type. Although this work was made on silicon-like systems with the use of a model Stillinger-Weber interaction potential, the overdamped dynamics enlarges its domain of applicability to a qualitative study of foams and colloidal systems, as attested by good experimental comparisons~\cite{b.mobius,b.schall1}. Similar approaches could be used to infer the different relaxation times observed in the experiments, for example in gels~\cite{b.divoux}. This work opens also new perspectives to understand the dependence of the size and of the number of plastic basins as a function of the chemical specificities of interatomic interactions, such as the bond directionality. In particular, it could be used to understand the respective role of the local structure and of the long-range mechanical interactions on the small scale plasticity of disordered materials.

\end{document}